%% file: vbf_hP2j.tex
\documentclass[11pt,a4paper,hyper]{JHEP3}
\usepackage{epsfig,amsmath,multirow}

\newlength{\largfig}
\def\beq{\begin{equation}} 
\def\eeq{\end{equation}} 
 
\def\beqn{\begin{eqnarray}} 
\def\eeqn{\end{eqnarray}}

\def\hjj{Hjj}

\def\eps{\epsilon}
\def\mc{\mathcal}
\def\mr{\mathrm}

\newcommand{\bea}{\begin{eqnarray}}
\newcommand{\eea}{\end{eqnarray}}
\def\beq{\begin{equation}}
\def\eeq{\end{equation}}
\def\bec{\begin{center}}
\def\eec{\end{center}}

\newcommand{\non}{\nonumber}
\def\mun{\mu_0}
\def\muf{\mu_\mr{F}}
\def\mur{\mu_\mr{R}}
\def\xif{\xi_\mr{F}}
\def\xir{\xi_\mr{R}}
\def\eps{\varepsilon}
\def\als{\alpha_s}

\def\hjj{H j j }
\def\hajj{H \gamma j j }
\def\MB{{\cal M}_B}
\def\MV{{\cal M}_V}

\def\lq{\left[} 
\def\rq{\right]} 
 \def\kfac{\mr{K~factor}}
 \def\kf{\mr{K}}
 \def\qsc{Q_i^2}
 \def\mptsc{m_H^2+\sum p_{Tj}^2}

\newcount\minutes 
\newcount\scratch 
 
\def\timestamp{%
\scratch=\time 
\divide\scratch by 60 
\edef\hours{\the\scratch} 
\multiply\scratch by 60 
\minutes=\time 
\advance\minutes by -\scratch 
---$\,$\hours:\null 
\ifnum\minutes< 10 0\fi 
\the\minutes} 

\title{Next-to-leading order QCD corrections to Higgs boson production in 
association with a photon via weak-boson fusion at the LHC}

\author{Ken Arnold \\
Institute for Theoretical Physics, Karlsruhe Institute of Technology, \\ 76128~Karlsruhe, Germany \\
E-mail: \email{arnold@particle.uni-karlsruhe.de}}
\author{Terrance Figy \\
CERN, CH--1211 Geneva 23, Switzerland \\
E-mail: \email{terrance.maynard.figy@cern.ch}}
\author{Barbara J\"ager \\
Institute for Theoretical Physics and Astrophysics, University of W\"urzburg, 97074~W\"urzburg, Germany \\
E-mail: \email{bjaeger@physik.uni-wuerzburg.de}}
\author{Dieter Zeppenfeld \\
Institute for Theoretical Physics, Karlsruhe Institute of Technology, \\ 76128~Karlsruhe, Germany \\
E-mail: \email{dieter@particle.uni-karlsruhe.de}}
\abstract{
Higgs boson production in association with a hard central photon and two forward tagging jets is expected to provide valuable information on Higgs boson couplings in a range where it is difficult to disentangle weak-boson fusion processes from large QCD backgrounds. 
We present next-to-leading order QCD corrections to Higgs production in association with a photon via weak-boson fusion at a hadron collider in the form of a flexible parton-level Monte Carlo program.  
The QCD corrections to integrated cross sections are found to be small for experimentally relevant selection cuts, while the shape of kinematic distributions can be distorted by up to 20\% in some regions of phase space. Residual scale uncertainties at next-to-leading order are at the few-percent level. 
}


\preprint{CERN-PH-TH/2010-119}
\begin{document}

%
\input{intro.tex}
\input{details.tex}

\input{results.tex}

\input{conclusions.tex}
%
%
%
\acknowledgments{
We would like to thank F.~Campanario for useful discussions. 
This work was supported in part by the Deutsche Forschungsgemeinschaft
under SFB TR-9 ``Computational Particle Physics'' and via the 
Graduiertenkolleg ``High Energy Physics and Particle Astrophysics'', and by the Initiative and Networking Fund of the Helmholtz Association, contract HA-101 ("Physics at the Terascale"). 
T.~F. would like to thank the Institute for Particle Physics Phenomenology at Durham University and the CERN Theory Division for their support. }
%
%
\bibliographystyle{JHEP}
\bibliography{ref}

\end{document}

%% file: intro.tex
\section{Introduction}
\label{sec:intro}
The start-up of the CERN Large Hadron Collider (LHC) marks a new era of high energy particle physics. A major goal of the LHC is the discovery of the Standard Model (SM) Higgs boson and the determination of its properties~\cite{:1999fr,Ball:2007zza}. In this context, weak-boson fusion (WBF) processes have been identified as an important class of reactions. In particular, Higgs production via WBF, i.e.\ the electroweak (EW) reaction $qq\to qqH$, where the decay products of the Higgs boson in the central-rapidity range are detected in association with two tagging jets of large invariant mass, provides a promising discovery channel for the Higgs boson~\cite{Rainwater:1999sd,Kauer:2000hi,Rainwater:1998kj,Rainwater:1997dg}. 
Once the Higgs boson has been found and its mass determined, WBF will allow for a determination of its CP properties~\cite{Plehn:2001nj,Hankele:2006ma} and couplings to gauge bosons and fermions~\cite{Zeppenfeld:2000td,
Duhrssen:2004cv}. Combining information from the $H\to\tau\tau$, $H\to W^+W^-$,  $H\to\gamma\gamma$, and $H\to\mr{invisible}$ channels, the couplings of the Higgs boson to the top quark, tau lepton, and the weak gauge bosons can be constrained with an accuracy dictated by the amount of data available. Because of challenging requirements on the {\tt ATLAS} and {\tt CMS} triggers and large QCD backgrounds, however, the determination of the $Hb\bar b$ coupling in Higgs production via WBF remains difficult~\cite{Mangano:2002wn}.  
Therefore, new search strategies have been suggested, such as making use of the sub-structure of so-called ``fat jets'', resulting from the bottom quarks into which a boosted Higgs boson decays in $WH$ and $ZH$~\cite{Butterworth:2008iy} or in $t\bar t H$ production~\cite{Plehn:2009rk} at the LHC.  
Additional constraints on the bottom quark Yukawa coupling could be provided by a future high-energy lepton-hadron collider such as the CERN Large Hadron electron Collider (LHeC), which offers a cleaner environment than a hadron-hadron collider~\cite{Dainton:2006wd,Han:2009pe,Blumlein:1992eh,Jager:2010zm}. 

Alternatively, extra gauge boson radiation in WBF can serve as a valuable tool for improving the signal-to-background (S/B) ratio of the $Hb\bar b$ mode in a  hadron-collider environment.
In Ref.~\cite{Rainwater:2000fm}, $WH$ production via WBF has been found to allow for distinguishing a $H b\bar b$ coupling compatible with the SM from certain other scenarios.  The event statistics of the WBF $WH$ mode is limited, however, by the requirement of a leptonic decay of the $W$ boson~\cite{Rainwater:2000fm,Ballestrero:2008kv}. This loss in statistics can be avoided by requesting a hard photon rather than a massive gauge boson in association with the Higgs boson produced via WBF~\cite{Gabrielli:2007wf,Asner:2010ve}. Indeed, it has been shown that the presence of a central photon can improve triggering efficiencies for the multi-jet final state needed to select $pp\to H(\to b\bar b)jj$ events. Moreover, due to a large gluonic component, QCD backgrounds to the $b\bar b jj$ final state are less active in radiating a photon of large transverse momentum than the quark-dominated WBF signal. Additional interference effects have been found to suppress backgrounds even further~\cite{Gabrielli:2007wf}. The consequence of the central photon requirement is thus a pronounced increase in the S/B ratio, making the channel $pp\to \hajj$ a process particularly worthwhile to investigate.  

In view of the importance of this channel, precise predictions for the signal process are essential. 
We provide next-to-leading order QCD corrections to $\hajj$ production via WBF at a hadron collider in the form of a parton-level Monte Carlo program, structured similarly to existing code for WBF-type reactions~\cite{Arnold:2008rz}. The program allows for the calculation of cross sections and distributions within experimentally relevant selection cuts. In order to obtain infrared-safe predictions for the $\hajj$ final state, we employ the photon-isolation criterion of Frixione~\cite{Frixione:1998jh}. 

We start with a brief description of the calculation in Sec.~\ref{sec:calc}. In Sec.~\ref{sec:num} we provide a detailed phenomenological study of $\hajj$ production via WBF at NLO-QCD accuracy. We estimate the theoretical uncertainties of our predictions by analyzing the scale dependences of integrated cross sections and present kinematic distributions within different experimental settings. The impact of the QCD corrections on various observables is quantified. Our conclusions are given in Sec.~\ref{sec:conc}.  

%% file: details.tex
\section{Details of the calculation}
\label{sec:calc}
\subsection{Tree-level calculation and approximations}
\begin{figure}
\bec
\includegraphics[width=0.9\textwidth,clip]{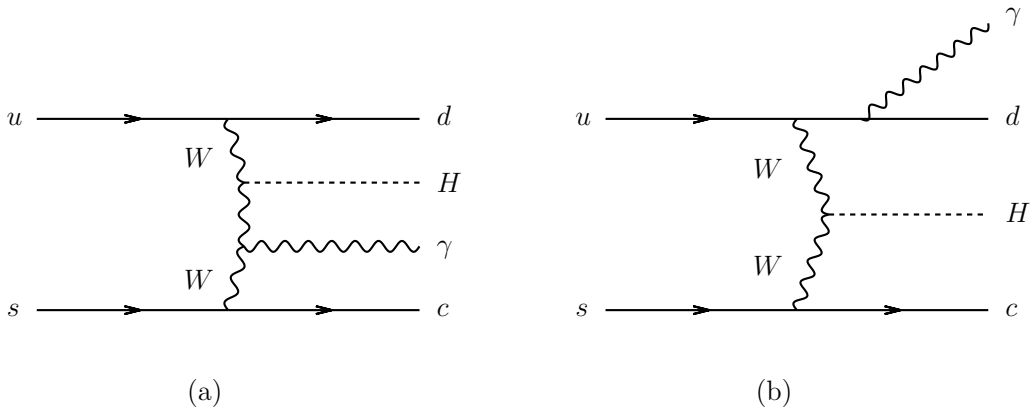}
\caption{
Feynman diagrams contributing to the Born process $us\to dcH\gamma$. Graphs analogous to~(b), with the photon being emitted off the lower quark line,  are not shown.}
\label{fig:born}
\eec
\end{figure}

At proton-proton colliders, Higgs boson production in association with a photon in WBF mainly proceeds via quark-quark scattering processes, $qq'\to qq'H\gamma$, mediated by the exchange of a $W^\pm$ or a $Z$~boson in the $t$-channel. The Higgs boson is radiated off this weak boson, while the photon can be emitted either from a fermion line or from a $t$-channel exchange $W$ boson. The relevant charged-current (CC) Feynman diagrams can thus be grouped in two topologies, depending on how many gauge bosons couple to a fermion line. Representative graphs for each topology of a specific subprocess are depicted in Fig.~\ref{fig:born}. 
%
%
%
%
Since, within the SM, the photon does not couple directly to $Z$ bosons, only graphs corresponding to topology~(b) contribute to neutral current (NC) production modes. 

For the calculation of the matrix elements we employ the methods applied already to a variety of WBF reactions~\cite{Jager:2006cp,Jager:2006zc,Bozzi:2007ur,Jager:2009xx,Figy:2007kv,Figy:2008zd}, based on the helicity-amplitude techniques of Ref.~\cite{Hagiwara:1985yu,Hagiwara:1988pp}. 
We decompose all Feynman diagrams into fermionic currents for each quark line and bosonic tensors parameterizing the gauge-boson interaction in the $t$-channel. For the graphs of topology~(a), quark currents for both fermion lines and a bosonic tensor for the $W^+ W^-\to H\gamma$ sub-amplitude are needed. Topology~(b) furthermore requires a quark current including photon emission and bosonic tensors for the sub-amplitude $W^+ W^-\to H$ in CC processes and  $ZZ\to H$ in NC modes. We have developed three different implementations of the tree-level matrix elements, which are supplemented by code adapted from (i)~{\tt FeynArts}~\cite{feynarts} and~{\tt Formcalc} \cite{formcalc} and (ii,iii)~{\tt HELAS}~\cite{Murayama:1992gi}.  
In each implementation, building blocks entering in various diagrams are stored and evaluated once only. 
Diagrams for the related processes $q\bar q'\to q\bar q'H\gamma$, $\bar qq'\to \bar qq'H\gamma$, and  $\bar q\bar q'\to \bar q\bar q'H\gamma$ are easily obtained by crossing. 

In addition to the topologies discussed above, annihilation processes such as $q\bar q\to ZH\gamma$ with subsequent decay $Z\to q\bar q$ and similar $WH\gamma$ production modes occur. In sub-processes with identical quarks, interference contributions of $t$-channel with $u$-channel diagrams can arise. In the phase space regions where WBF processes can be observed experimentally, however, with widely separated quark jets of large invariant mass, these types of  contributions are entirely negligible~\cite{Ciccolini:2007ec}. We therefore disregard them throughout. For simplicity, we refer to $pp\to\hajj$ within the approximations discussed as ``EW $\hajj$'' production.

\subsection{Calculation of the NLO-QCD corrections}
The calculation of real-emission corrections to EW $\hajj$ production proceeds along the same lines as the LO computation. A gluon has to be attached to the fermions in all possible ways, yielding $qq'\to qq'gH\gamma$ contributions and related sub-processes with anti-quarks. Crossing the gluon to the initial state gives rise to (anti-)quark-gluon initiated modes with an additional fermion in the final state, such as $gq'\to q\bar q q'H\gamma$. The form of the bosonic tensors that already occurred at LO is not affected. 
Singularities in soft and collinear regions of phase space are handled  in two different regularization schemes: in conventional dimensional regularization and in dimensional reduction, with $d=4-2\eps$ space-time dimensions. The cancellation of these divergences with respective poles in the virtual corrections is performed by introducing the counter-terms of the dipole subtraction formalism of Ref.~\cite{Catani:1996vz}. Since the QCD structure of EW~$\hajj$ production is identical to the related case of $pp\to\hjj$ via WBF, the counter-terms are of the same form and can straightforwardly be adapted from Ref.~\cite{Figy:2003nv}, where also explicit expressions for the finite collinear terms are given.  

The virtual corrections comprise the interference of one-loop diagrams with the Born amplitude. Due to color conservation, only selfenergy, vertex, and box corrections to either the upper or the lower quark line need to be considered. Contributions from graphs with a gluon being attached to both the upper and the lower quark lines vanish at order $\als$, within our approximations. 
The interference of the relevant diagrams containing one-loop corrections to either the upper or the lower quark line, $\MV^{(i)}$, with the Born amplitude, $\MB$, is of the form
\bea
\label{eq:virtual_born}
2\,\mr{Re} \lq \MV^{(i)}\MB^* \rq
&=& |\MB|^2 \frac{\alpha_s(\mur)}{2\pi} C_F
\left(\frac{4\pi\mur^2}{Q_i^2}\right)^\epsilon \Gamma(1+\epsilon)\\ \non
 &&\times
\lq-\frac{2}{\epsilon^2}-\frac{3}{\epsilon}+c_{\rm virt}\rq\
+2\,\mr{Re} \lq \widetilde{\MV}^{(i)}\MB^* \rq \,,
\eea
where  $Q_i$ is related to the momentum transfer between the respective initial- and the final-state quarks carrying the momenta $k_1$ and $k_2$ via $Q_i^2 = -(k_1-k_2)^2$, $\mu_R$ is the renormalization scale, $C_F=4/3$, and $c_{\rm virt}$ is a constant, given by $c_{\rm virt}=\pi^2/3-8$ in conventional dimensional regularization and by $c_{\rm virt}=\pi^2/3-7$ in dimensional reduction. The quantity $\widetilde{\MV}^{(i)}$ is a finite remainder. 

In order to compute the $\widetilde{\MV}^{(i)}$, we have split off the divergent pieces and expressed the remainder in terms of the finite parts of the Passarino-Veltman $B_0$, $C_{jk}$, and $D_{jk}$ functions~\cite{Passarino:1978jh}, which are evaluated numerically. 
To this end, we have prepared three different implementations, which are based on the Passarino-Veltman tensor reduction procedures of Ref.~\cite{Figy:2007kv}, Ref.~\cite{Campanario:2008yg,Campanario:2009um,Bozzi:2009ig,Campanario:2010hp}, and Refs.~\cite{Bredenstein:2008tm,Jager:2010aj}, respectively.  
The tensor-reduction procedure may give rise to numerical instabilities for certain phase-space points, due to small Gram determinants emerging in the determination of the box-type corrections. We monitor these numerical instabilities carefully by requiring electroweak Ward-identities for all box-type contributions to be fulfilled with an accuracy of $10^{-3}$ or better. We find that less than 0.1~permille of the generated events fails to meet this condition. The problematic contributions from these phase-space points to the finite parts of the box diagrams are negligible and therefore disregarded.  

The poles in Eq.~(\ref{eq:virtual_born}) are canceled by respective singularities in the phase-space integrated counter-terms, which in the notation of Ref.~\cite{Catani:1996vz} are given by
\beq
\label{eq:I}
\langle \mc{I}(\epsilon)\rangle  = |\MB|^2 \frac{\alpha_s(\mur)}{2\pi} C_F
\left(\frac{4\pi\mur^2}{Q_i^2}\right)^\epsilon \Gamma(1+\epsilon)
\lq\frac{2}{\epsilon^2}+\frac{3}{\epsilon}+9-\frac{4}{3}\pi^2\rq\;.
\eeq

\subsection{Checks}
All building blocks entering the LO and NLO cross sections for EW~$\hajj$ production have been tested extensively. As mentioned above, we have prepared three different implementations of the tree-level, real emission, and virtual matrix elements. We found perfect agreement between these implementations at the amplitude level. The tree-level and real-emission contributions have also been compared to fully automatically generated amplitudes provided by {\tt MadGraph}~\cite{Stelzer:1994ta}. The matrix elements agree to about 12 digits for a representative set of phase-space points. 

Furthermore, we have verified the QCD gauge invariance of the real-emission contributions, and the QED gauge invariance of the tree-level, virtual, and real-emission amplitudes by checking that they vanish upon replacing the polarization vector of the gluon and of the photon, respectively, with the corresponding momenta.  

In addition, we compared integrated cross sections at NLO-QCD accuracy within various settings, including those of Sec.~\ref{sec:num}. All results agreed within the respective relative statistical errors, which are at the level of less than $10^{-3}$ for Monte-Carlo runs with samples of about 50~million phase-space points. 
Integrated LO cross sections within inclusive selection cuts have also been compared to the corresponding results of {\tt MadEvent}~\cite{Maltoni:2002qb,Alwall:2007st}. The cross sections agree within the accuracy of the two programs. 

A precise comparison of LO cross sections with those of Ref.~\cite{Gabrielli:2007wf} was difficult, since not all EW parameters of the calculation have been listed in the publication. We have thus used our ``default'' settings of Sec.~\ref{sec:num} for all parameters not explicitly listed in Ref.~\cite{Gabrielli:2007wf}. With this prescription, we could reproduce their cross sections with an accuracy of $3-5\%$, dependent on the chosen cuts.

%% file: results.tex
\section{Numerical results}
\label{sec:num}

For our numerical analysis, we assume a hadronic center-of-mass~(c.m.)~energy of $\sqrt{S}=14$~TeV, unless stated otherwise. We use the CTEQ6M parton distribution functions with $\als(m_Z)=0.118$ at NLO, and the CTEQ6L1 set at LO~\cite{Pumplin:2002vw} as a default. We chose $m_Z=91.188$~GeV, $m_W=80.398$~GeV, and the measured value of $G_F=1.166\times 10^{-5}/$GeV$^2$ as electroweak input parameters, from which $\alpha_\mr{QED}$ and $\sin^2\theta_W$ are computed via LO electroweak relations. 
Throughout our calculation, fermion masses are set to zero, and contributions with external top-quarks are disregarded. Subprocesses with $b$ quarks in the initial state are considered for the NC modes, while CC processes comprising the transition of a $b$ to a $t$ quark are not taken into account.  
For the Cabibbo-Kobayashi-Maskawa matrix, $V_\mr{CKM}$, we have used a diagonal form, equal to the identity matrix, which is equivalent to employing the exact $V_\mr{CKM}$ when the summation over final-state quark flavors is performed and quark masses are neglected. 

In order to reconstruct jets from the final-state partons,
the $k_T$ algorithm~\cite{Catani:1992zp,Catani:1993hr,Ellis:1993tq} as described in Ref.~\cite{Blazey:2000qt} is used, with resolution parameter $D=0.7$. 
Jets are required to have 
\begin{eqnarray}
\label{eq:cuts-jet}
p_{Tj} \geq 20~{\rm GeV} \, , \qquad\qquad |y_j| \leq 5 \, .
\end{eqnarray}
Here $p_{Tj}$ denotes the transverse momentum of a jet, and $y_j$ the rapidity of the (massive) jet momentum which is reconstructed as the four-vector sum of massless partons of pseudo-rapidity $|\eta_j|<5$. 
At LO, there are exactly two massless final state partons, which are identified as tagging jets, provided they pass the $k_T$ algorithm and the cuts described above. At NLO, a third parton may be encountered which can either be recombined with another parton or give rise to an additional jet. In this case, we choose to identify the two jets of highest transverse momentum as ``tagging jets''. 

For the Higgs boson, we simulate a generic decay into two massless particles without specifying a particular channel. The decay particles,  each one labeled $d$, can represent, for instance, $b\bar b$ final states. The respective branching ratio, $BR(H\to dd)$, is not included in the numerical results presented below. 
The photon is isolated in a theoretically well-defined way with the help of the criterion suggested in Ref.~\cite{Frixione:1998jh}, which allows us to avoid introducing parton-to-photon fragmentation contributions. An event is considered as acceptable, if the hadronic energy deposited in a cone around the direction of the photon is limited by 
\beq
\label{eq:acone}
\sum_{i: \Delta R_{i\gamma}<\Delta R}
      p_{Ti} \leq \frac{1-\cos \Delta R}{1-\cos\delta_0} p_{T\gamma} 
\qquad
(\forall\Delta R\leq \delta_0)\,. 
\eeq
Here, the summation index $i$ runs over all final-state partons found in a cone of size $\Delta R$ in the rapidity-azimuthal angle plane around the photon, $p_{Ti}$ denotes the transverse momentum, and $\Delta R_{i\gamma}$ the separation of parton $i$ from the photon, while $\delta_0$ stands for a fixed separation. 

In order to explore the impact of NLO-QCD corrections in different regimes, we have performed phenomenological analyses with two sets of selection cuts in addition to the common jet-defining criteria of Eq.~(\ref{eq:cuts-jet}), to which we refer as ``inclusive cuts'' and ``WBF cuts'', respectively. 

For the ``inclusive cuts'' scenario, the Higgs decay particles are required to be separated from each other and from the other final-state particles by
\beq
\label{eq:icuts-rik}
\Delta R_{dd} > 0.4\;, \quad
\Delta R_{d\gamma} > 0.4\;, \quad
\Delta R_{jd} > 0.4\;, \quad
\Delta R_{j \gamma} > 0.4 \;,
\eeq  
where each $\Delta R_{kk'}$ ($\Delta R_{jk'}$) denotes the separation of particle $k$ (jet $j$) from particle $k'$ in the rapidity-azimuthal angle plane. 

The cone-size parameter of Eq.~(\ref{eq:acone}) is set to
\beq
\delta_0 = 0.7\;.
\eeq
The decay particles and the photon need to be located in the central-rapidity range of the detector, 
\beq
\label{eq:cuts-yad}
|y_d| < 2.5\;,\quad 
|y_\gamma| < 2.5\;,
\eeq
and exhibit sufficiently large transverse momenta, 
\beq
\label{eq:cuts-ptad}
p_{T d} > 20~\mr{GeV}\;, \quad
p_{T \gamma} > 20~\mr{GeV}\;.
\eeq
The two tagging jets are required to fulfill the invariant mass criterion of
\beq
\label{eq:icuts-mjj}
M_{jj}^\mr{tag}>100~\mr{GeV}\,.
\eeq

More stringent constraints are applied for the ``WBF cuts'' scenario. In addition to the cuts of Eqs.~(\ref{eq:cuts-jet}), (\ref{eq:cuts-yad}), and (\ref{eq:cuts-ptad}) we now require 
\beq
\label{eq:cuts-rik}
\Delta R_{dd} > 0.7\;, \quad
\Delta R_{d\gamma} > 0.7\;, \quad
\Delta R_{jd} > 0.7\;,\quad 
\Delta R_{j \gamma} > 0.7\;, \quad
\delta_0 = 0.7\;.
\eeq  
Backgrounds to WBF are significantly suppressed by imposing a large rapidity separation on the two tagging jets, 
\beq
\label{eq:cuts-yjj}
\Delta y_{jj} = |y_{j1} - y_{j2}| > 4\;,
\eeq
with the photon and the decay products of the Higgs boson being located in between the tagging jets,
\beq
\min(y_{j1},y_{j2}) \leq y_\gamma, y_d \leq \max(y_{j1},y_{j2})\,.
\eeq 
The tagging jets are furthermore required to reside in opposite detector hemispheres with
\beq
y_{j1}\times y_{j2}<0\,,
\eeq
and exhibit a large invariant mass, 
\beq
\label{eq:cuts-mjj}
M_{jj}^\mr{tag}>600~\mr{GeV}\,.
\eeq

\begin{figure}[tp!]
\begin{center}
\includegraphics[scale=0.8]{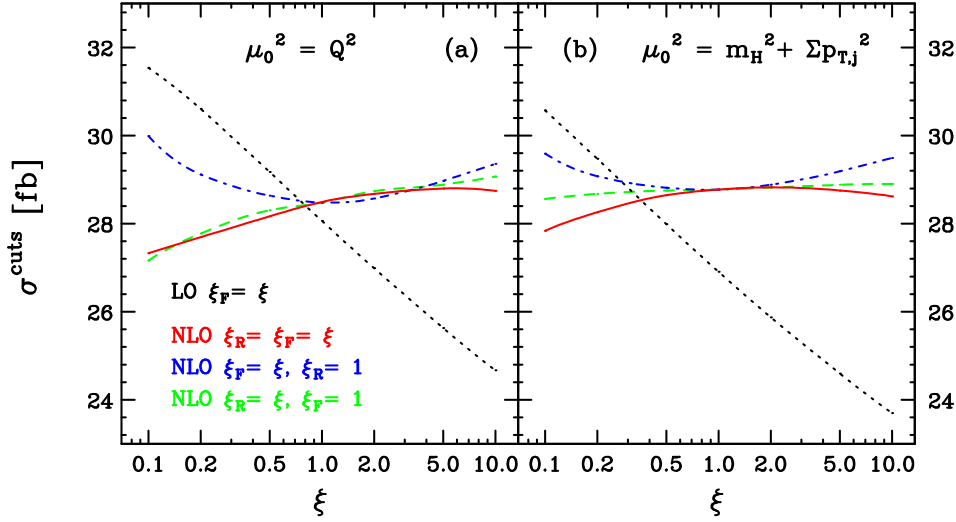}
\caption{
Scale dependence of the integrated cross section within the inclusive cuts of 
Eq.~(\protect\ref{eq:cuts-jet}) and Eqs.~(\protect\ref{eq:icuts-rik})--(\protect\ref{eq:icuts-mjj}) at LO 
and NLO for two different choices of $\mu_{0}$. Shown are curves for $\xif = \xi$ at LO (black dots), $\xir=\xif=\xi$ at NLO (red solid), $\xif=\xi$, $\xir=1$ at NLO (blue dash-dot), and $\xir=\xi$, $\xif=1$ at NLO (green dashes).
}
\label{fig:scale-dep}
\end{center}
\end{figure}

In order to estimate the dependence of our predictions on unphysical scales, we have computed the integrated cross section within the inclusive cuts of Eq.~(\ref{eq:cuts-jet}) and Eqs.~(\ref{eq:icuts-rik})--(\ref{eq:icuts-mjj}), $\sigma^\mr{cuts}$, for two different choices of the factorization and renormalization scales, $\muf$ and $\mur$, which are taken as multiples of the scale parameter $\mun$, 
\beq
\muf = \xif\,\mu_0\,,\qquad 
\mur = \xir\,\mu_0\,. 
\eeq
Figure~\ref{fig:scale-dep}~(a) shows our results for 
%
%
%
%
$\mun^{2} = \qsc$, where for each fermion line $\mu_0$ is determined as the momentum transfer carried by the virtual weak boson emitted from it.  
In Fig.~\ref{fig:scale-dep}~(b), $\sigma^\mr{cuts}$ is given for  $\mun^2=\mptsc$, where for each event the sum runs over the transverse momenta of all identified jets. Qualitatively, the results resemble the scale dependence of related WBF reactions \cite{Jager:2010aj}. 
In the range $1/2\leq\xif=\xir\leq 2$, the LO cross section decreases by about 8\% in each case, while the NLO prediction changes by about 2\% for $\mun^{2} = \qsc$ and less than 1\% for $\mun^2=\mptsc$. The $\kfac$, defined as the ratio of the LO cross section to the respective NLO result, is close to one for the former choice ($\kf=1.02$), while $\kf=1.07$ for $\mun^2=\mptsc$.

In order to quantify the impact of the parton distribution functions on the integrated cross section within the WBF-specific selection cuts of Eq.~(\ref{eq:cuts-jet}) and Eqs.~(\ref{eq:cuts-rik})--(\ref{eq:cuts-mjj}), $\sigma^\mr{WBF}$, in Table~\ref{tab:CTEQvsMSTW} we list the respective LO and NLO predictions as obtained with our default set, CTEQ6, and with the MSTW parton distributions of Ref.~\cite{Martin:2009iq} for the two scale settings discussed above. 
Whenever the MSTW set is used, the corresponding expression for the strong coupling is employed. 
%
%
\begin{table}[t!]
  \begin{center}
    \fbox{%
    \begin{tabular}{|c|c|c|c|c|}
      \multicolumn{5}{c}{\multirow{2}{*}{\bf $\boldsymbol{\sigma^\mr{WBF}[\text{fb}]$ for $\sqrt{S}=14}$ TeV}}\\
      \multicolumn{5}{c}{}\\
      \hline
      \multicolumn{5}{|c|}{LO}\\
      \hline
      \hline
      &\multicolumn{2}{c|}{CTEQ6}&\multicolumn{2}{c|}{MSTW}\\
      \hline
      $\xi$ & $\mun^{2} = \qsc$ & $\mun^2=\mptsc$ & $\mun^{2} = \qsc$ & $\mun^2=\mptsc$\\
      \hline
      $0.5$ & $15.72$ & $ 14.56 $ & $15.53  $ & $14.30$\\
      \hline
      $1.0$ & $14.65$ & $ 13.61 $ & $14.40  $ & $13.30$\\
      \hline
      $2.0$ & $13.70$ & $ 12.76 $ & $13.40  $ & $12.42$\\
      \hline
      \multicolumn{5}{c}{}\\
      \hline
      \multicolumn{5}{|c|}{NLO}\\
      \hline
      \hline
      &\multicolumn{2}{c|}{CTEQ6}&\multicolumn{2}{c|}{MSTW}\\
      \hline
      $\xi$ & $\mun^{2} = \qsc$ & $\mun^2=\mptsc$ & $\mun^{2} = \qsc$ & $\mun^2=\mptsc$\\
      \hline
      $0.5$ & $14.60$ & $14.84  $ & $14.70  $ & $14.93 $\\
      \hline
      $1.0$ & $14.79$ & $14.84  $ & $14.91  $ & $14.95 $\\
      \hline
      $2.0$ & $14.83$ & $14.75  $ & $14.94  $ & $14.85$\\
      \hline
    \end{tabular}%
  }
    \vspace{2\baselineskip}
    \caption{Cross sections obtained for different values of the scale factor $\xi=\xi_F=\xi_R$ within the ``WBF cuts'' scenario of Eq.~(\protect\ref{eq:cuts-jet}) and Eqs.~(\protect\ref{eq:cuts-rik})--(\protect\ref{eq:cuts-mjj}). The relative statistical errors of the quoted results are at the sub-permille level.}
  \label{tab:CTEQvsMSTW}
  \end{center}
\end{table}
%
%
The differences between the LO cross sections for different parton distribution functions (but apart from that identical settings) are at the level of 2\% and thus much smaller than those caused by the choice of the factorization scale. For instance, $\sigma^\mr{WBF}_\mr{LO}(\muf^2=\qsc)$ and $\sigma^\mr{WBF}_\mr{LO}(\muf^2=\mptsc)$ differ by more than 7\% for CTEQ6 and 8\% for MSTW. In analogy to the ``inclusive cuts'' scenario discussed above, the scale dependence of $\sigma^\mr{WBF}$ is mitigated by the inclusion of NLO-QCD corrections. For the NLO cross sections, the scale uncertainty is small and comparable in size to the uncertainty due to the parameterization of the parton distributions functions.
In the following, we will use CTEQ6 parton distributions and set $\mun^{2} = \qsc$,  unless stated otherwise. 

\begin{figure}[tp!]
\begin{center}
\includegraphics[scale=0.8]{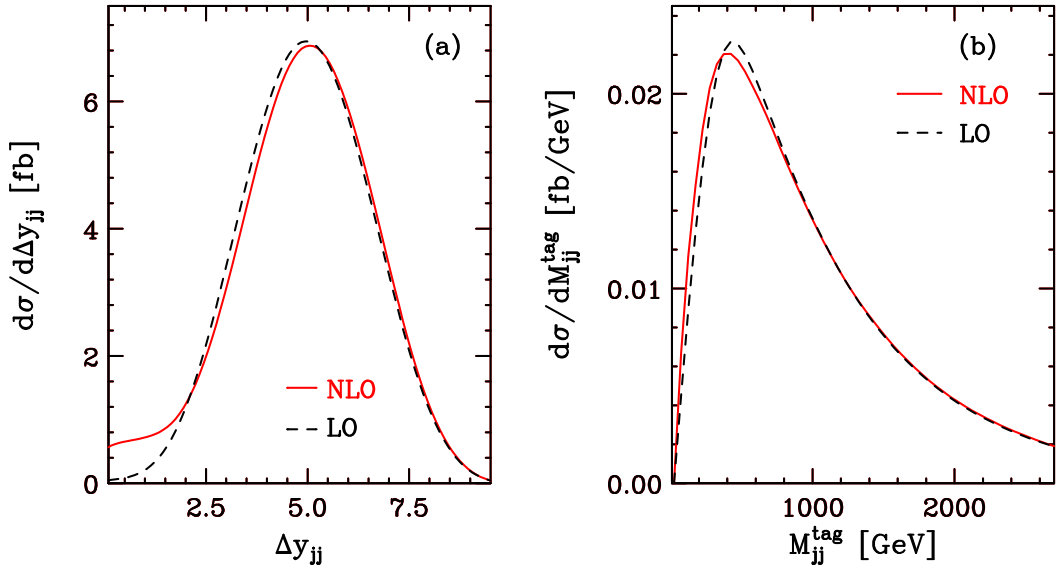}
\caption{Rapidity separation [panel~(a)] and invariant mass distribution of the two tagging jets [panel~(b)] in EW $\hajj$ production at the LHC with $\sqrt{S}=14$~TeV at LO (dashed black lines) and NLO (solid red lines), after the cuts of Eq.~(\protect\ref{eq:cuts-jet}) and Eqs.~(\protect\ref{eq:icuts-rik})--(\protect\ref{eq:icuts-mjj}) are applied. 
}
\label{fig:yjj-mjj}
\end{center}
\end{figure}

WBF-type reactions are characterized by widely separated hard jets in the far-forward and backward regions of the detector, being reflected by a large rapidity separation and invariant mass of the tagging jets. 
Figure~\ref{fig:yjj-mjj}~(a) illustrates the rapidity separation of the two tagging jets for the EW $\hajj$ cross section within the inclusive cuts of Eq.~(\ref{eq:cuts-jet}) and Eqs.~(\ref{eq:icuts-rik})--(\ref{eq:icuts-mjj}). 
%
%
%
%
Similar to the case of $Hjj$~\cite{Figy:2003nv,Berger:2004pca} and $Hjjj$~\cite{Figy:2007kv} production via WBF, the NLO-QCD corrections shift the peak of $d\sigma/d\Delta y_{jj}$ to slightly larger values. Due to the possible presence of a third jet in the real-emission contributions, at NLO an enhancement of events with small values of $\Delta y_{jj}$ occurs. Such contributions can be efficiently removed by imposing the rapidity-separation criterion of Eq.~(\ref{eq:cuts-yjj}).  
The shape of the invariant mass distribution, depicted in Fig.~\ref{fig:yjj-mjj}~(b), is barely affected by the NLO-QCD corrections. Since  $d\sigma/dM_{jj}^\mr{tag}$ peaks at rather large invariant masses and the distribution falls rather slowly at higher values of $M_{jj}^\mr{tag}$, the additional stringent cut of Eq.~(\ref{eq:cuts-mjj}) is powerful in suppressing QCD backgrounds which exhibit invariant mass distributions with a much steeper slope than the WBF-type signal process.  In the following we will therefore adhere to the WBF-specific cuts of Eqs.~(\ref{eq:cuts-rik})--(\ref{eq:cuts-mjj}) in addition to the generic requirements of Eqs.~(\ref{eq:cuts-jet}),~(\ref{eq:cuts-yad}),~and~(\ref{eq:cuts-ptad}). 

\begin{figure}[tp!]
\begin{center}
\includegraphics[scale=0.8]{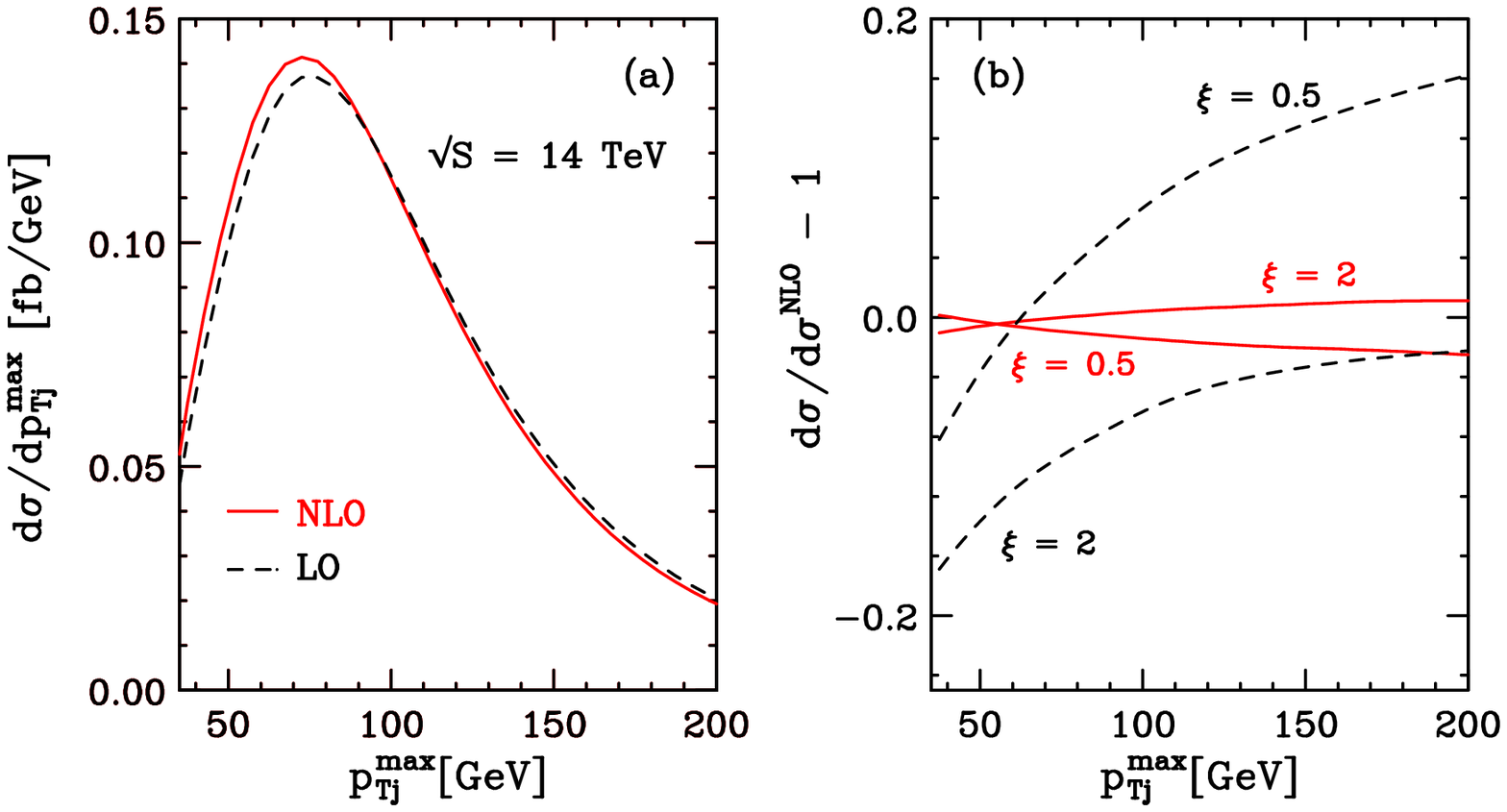}
\\
\vspace*{.5cm}
\includegraphics[scale=0.8]{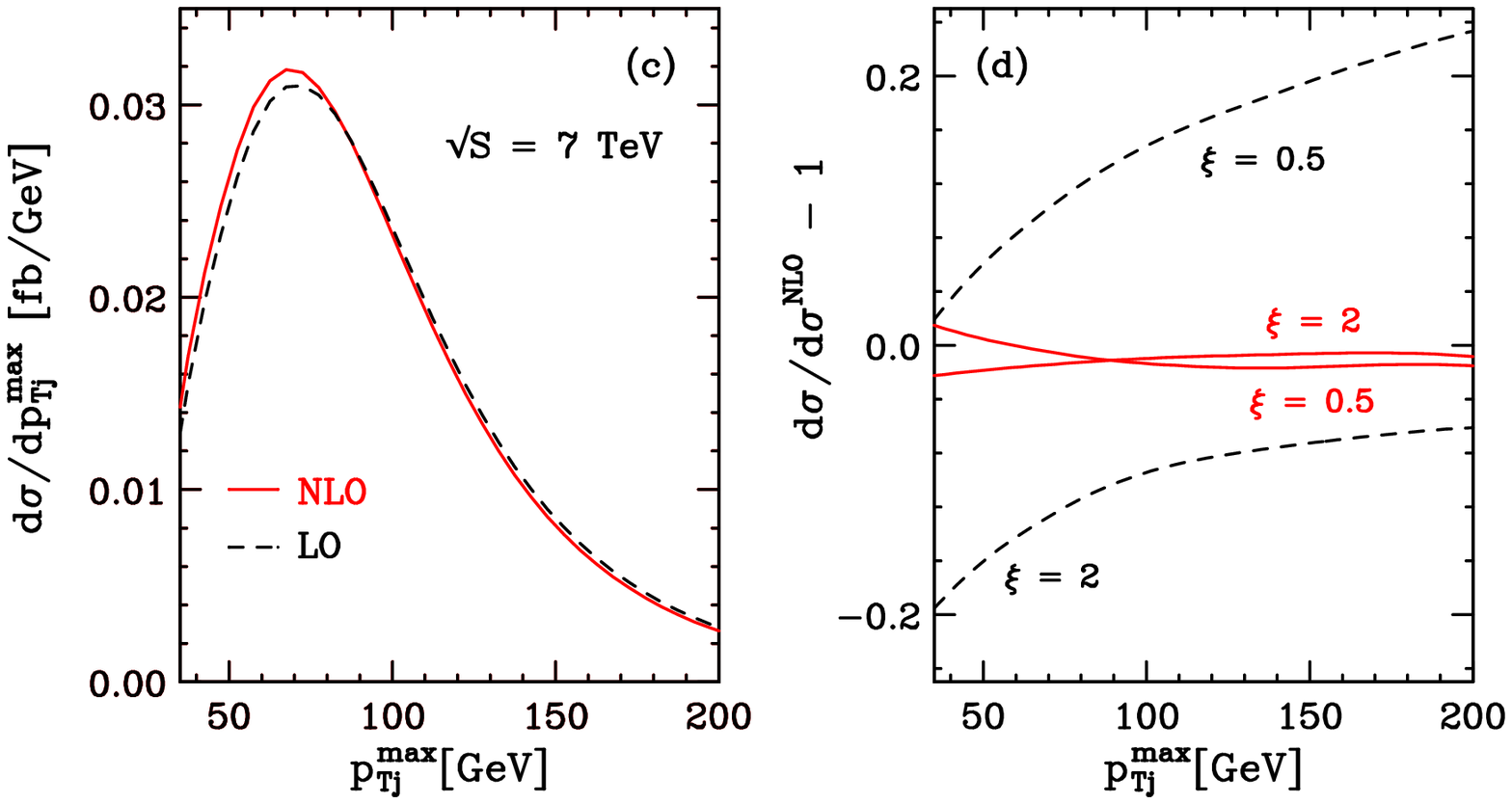}
\caption{Transverse momentum distribution of the hardest tagging jet in EW $\hajj$ production at the LHC with $\sqrt{S}=14$~TeV and $\sqrt{S}=7$~TeV, respectively, at LO (dashed black line) and NLO (solid red line) [panels~(a) and~(c)] and relative corrections according to Eq.~(\protect\ref{eq:delta}) when the factorization and renormalization scales are varied in the range $Q_i/2\leq\mur=\muf\leq 2Q_i$ [panels~(b)~and~(d)].  }
\label{fig:pttag}
\end{center}
\end{figure}

For this setting, the distribution of the hardest tagging jet is depicted in Fig.~\ref{fig:pttag}~(a). 
%
%
%
%
In order to assess the impact of the NLO-QCD corrections on the distribution of an observable~$\mc{O}$,  $d\sigma/d\mc{O}$, together with the scale uncertainties of the LO and the NLO prediction, we consider the quantity $\delta(\mc{O})$, defined as 
\beq
\delta(\mc{O}) = \frac{d\sigma(\xif,\xir)/d\mc{O}}{d\sigma^\mr{NLO}(\xif=\xir=1)/d\mc{O}} - 1\,, 
\label{eq:delta}
\eeq
where $d\sigma(\xif,\xir)/d\mc{O}$ denotes the LO or NLO expression, evaluated for arbitrary values of the scale parameters $\xif$ and $\xir$. The choice of $\mun$ is identical for $d\sigma/d\mc{O}$ and $d\sigma^\mr{NLO}/d\mc{O}$. 
In Fig.~\ref{fig:pttag}~(b), $\delta(p_{Tj}^\mr{max})$ is shown for $\mun^2=\qsc$ and two different values of the scale parameters, $\xi=\xif=\xir=1/2$ and 2. The difference  between the curves for the two values of $\xi$ indicates the scale uncertainty of $d\sigma/dp_{Tj}^\mr{max}$ at LO~(dashed black lines) and NLO~(solid red lines), respectively. For low transverse momenta, the NLO-QCD corrections are positive and modify the LO results by more than 10\%. With increasing  $p_{Tj}$, the scale uncertainty of the LO prediction becomes large, amounting to about 19\% for $p_{Tj}=200$~GeV. The NLO prediction, on the other hand, is stable against scale variations over the entire transverse-momentum range considered, changing by less than 4\% when $\xi$ is varied from 1/2 to 2 even for $p_{Tj}=200$~GeV. 
Figure~\ref{fig:pttag}~(c) 
displays $d\sigma/dp_{Tj}^\mr{tag,max}$  for a hadronic c.m.~energy of  $\sqrt{S}=7$~TeV. While the size of the cross section obviously goes down with $\sqrt{S}$, the peak structure of the transverse momentum distribution is barely affected when the energy is decreased from 14~TeV to 7~TeV. The relative scale uncertainty of the LO prediction, illustrated by Fig.~\ref{fig:pttag}~(d), is significantly larger for a lower collision energy, however. 

\begin{figure}[tp!]
\begin{center}
\includegraphics[scale=0.8]{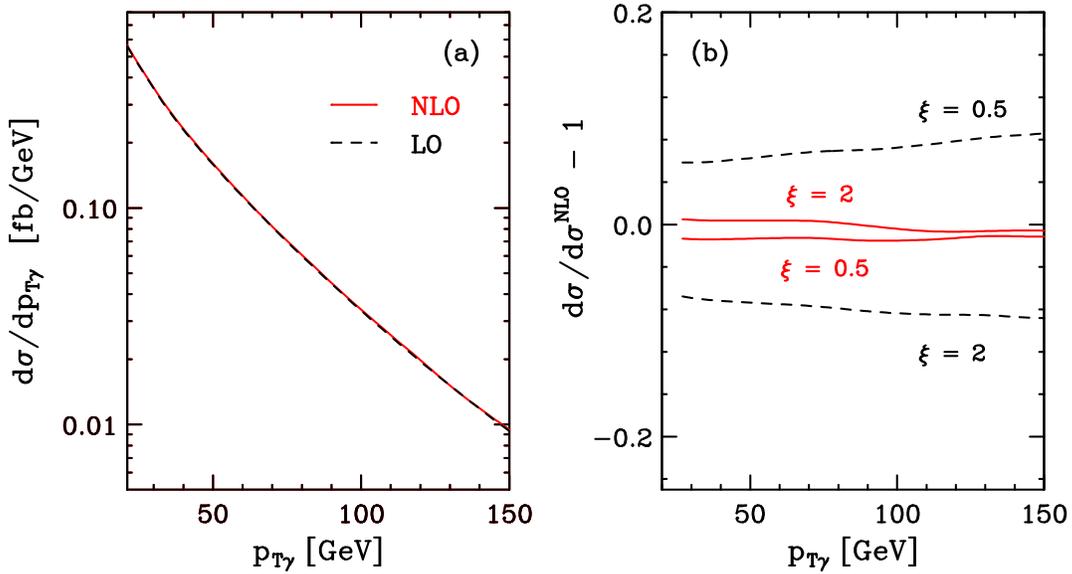}
\caption{Transverse momentum distribution of the photon in EW $\hajj$ production at the LHC with $\sqrt{S}=14$~TeV at LO (dashed black line) and NLO (solid red line) [panel~(a)] and relative corrections according to Eq.~(\protect\ref{eq:delta}) when the factorization and renormalization scales are varied in the range $Q_i/2\leq\mur=\muf\leq 2Q_i$ [panel~(b)].  }
\label{fig:pta}
\end{center}
\end{figure}

In contrast to the hardest tagging jet, the photon exhibits a transverse momentum distribution rather insensitive to NLO-QCD effects. For our default choice, $\mur^2=\muf^2=\qsc$, radiative corrections modify the LO result by less than 3\% over the entire range of $p_{T\gamma}$ considered. The residual scale variation  of the NLO-QCD prediction is very small. Our results for $d\sigma/dp_{T\gamma}$ and $\delta(p_{T\gamma})$ are shown in Fig.~\ref{fig:pta}.
%
%
%
%

\begin{figure}[tp!]
\begin{center}
\includegraphics[scale=0.8]{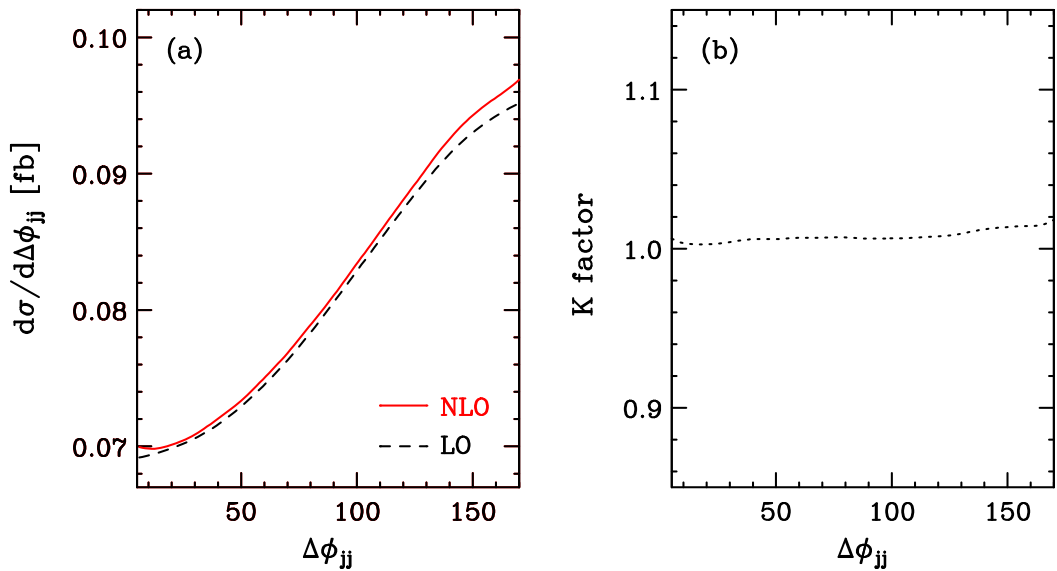}
\caption{Distribution of the azimuthal angle separation of the two tagging jets in EW $\hajj$ production at the LHC with $\sqrt{S}=14$~TeV at LO (dashed black line) and NLO (solid red line) [panel~(a)] and $\kfac$ according to Eq.~(\protect\ref{eq:kfac}) [panel~(b)].  
}
\label{fig:phijj}
\end{center}
\end{figure}

An observable particularly sensitive to the tensor structure of the Higgs coupling to weak bosons in WBF-type reactions is the azimuthal angle separation, $\Delta\phi_{jj}=|\phi_{j1}-\phi_{j2}|$, of the two tagging jets~\cite{Plehn:2001nj,Hankele:2006ma}. 
Figure~\ref{fig:phijj} 
%
%
%
%
illustrates $d\sigma/d\Delta\phi_{jj}$ for ``WBF cuts'' with $\mun^2=\qsc$ and $\sqrt{S}=14$~TeV together with the phase-space dependent $\kfac$, defined according to
\beq
\kf(\Delta\phi_{jj}) = \frac{d\sigma^\mr{NLO}(\muf,\mur)/d\Delta\phi_{jj}}{d\sigma^\mr{LO}(\muf,\mur)/d\Delta\phi_{jj}} \,, 
\label{eq:kfac}
\eeq
with the LO and the NLO distributions being evaluated for the same choice of scales. 
The shape of $d\sigma/d\Delta\phi_{jj}$ is rather insensitive to NLO-QCD corrections. 
Should azimuthal angle correlations very different from this prediction be observed in experiment, they could thus hint at coupling structures not accounted for within the SM.

\begin{figure}[tp!]
\begin{center}
\includegraphics[scale=0.8]{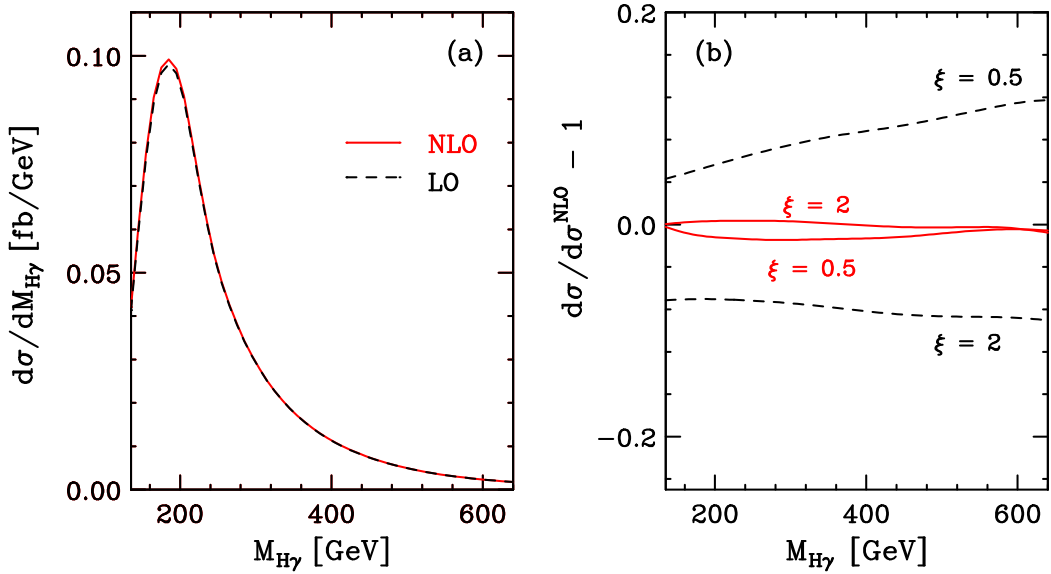}
\caption{Invariant mass distribution of the Higgs boson-plus-photon system in EW $\hajj$ production at the LHC with $\sqrt{S}=14$~TeV at LO (dashed black line) and NLO (solid red line) [panel~(a)] and relative corrections according to Eq.~(\protect\ref{eq:delta}) when the factorization and renormalization scales are varied in the range $Q_i/2\leq\mur=\muf\leq 2Q_i$ [panel~(b)].  }
\label{fig:mha}
\end{center}
\end{figure}

The invariant mass distribution of the Higgs boson-plus-photon system, being reconstructed from the four-momenta of the photon and the decay products of the Higgs boson, is shown in Fig.~\ref{fig:mha}  
%
%
%
%
for our default settings with $\mun^2=\qsc$ and $\sqrt{S}=14$~TeV. The distribution vanishes for $M_{H\gamma}<m_H$ and peaks for $m_H=120$~GeV at around $M_{H\gamma}\sim 165$~GeV. For larger Higgs masses, $d\sigma/dM_{H\gamma}$ is shifted to correspondingly higher values. Reducing $\sqrt{S}$ from 14~TeV to 7~TeV does not change the shape of the distribution, but increases the scale dependence of the LO prediction, in analogy to what has been observed above for the transverse momentum distribution of the tagging jets.

%% file: conclusions.tex
\section{Summary and conclusions}
\label{sec:conc}
In this work we have presented NLO-QCD corrections to Higgs production in association with a photon via weak-boson fusion at the LHC. We have developed a flexible parton-level Monte Carlo program which allows us to compute cross sections and kinematic distributions within experimentally relevant selection criteria, employing the photon-isolation procedure of Frixione~\cite{Frixione:1998jh}.   

We analyzed EW $\hajj$ production within two different settings. First, we imposed only minimal selection cuts to obtain a well-defined final-state configuration. Second, additional cuts were applied, designed to enhance WBF-type contributions with respect to QCD background processes. We found that in each case the impact of NLO-QCD corrections on integrated cross sections is small. The actual size of the $\kfac$ depends not only on the selection cuts, but also on the choice of the factorization and renormalization scales in the computation. We studied the two options $\mun^2=\qsc$ and  $\mun^2=\mptsc$, and found that slightly smaller NLO-QCD corrections are obtained for the former choice. The change of the NLO cross section when $\muf$ and $\mur$ are varied in the range $\mun/2\leq\muf=\mur\leq 2\mun$ is comparable in the two cases. 
NLO-QCD corrections do not only affect the overall normalization of the integrated cross sections, but also the shape of some kinematic distributions.   Relative corrections can be as large as 20\% in some regions of phase space. 